\documentclass[12pt,twoside]{article}
\usepackage{xrb2007}
\usepackage{epsfig}
\usepackage{graphicx}
\usepackage{longtable,lscape}

\def\integral{{\it INTEGRAL}}
\def\swift{{\it Swift}}
\def\xmm{{\it XMM-Newton}}
\def\chandra{{\it Chandra}}
\def\asca{{\it ASCA}}
\def\sax{{\it Beppo-SAX}}
\def \src {XMMU~J174716.1--281048}
\def \ax {AX~J1754.2--2754} 
\def \igr {IGR~J17464--2811}
\def \nh {$N_{\rm H}$}
\def \lerg {erg s$^{-1}$}
\def \ferg {erg cm$^{-2}$ s$^{-1}$}

\begin{document}

\session{Faint XRBs and Galactic LMXBs}

\shortauthor{Del Santo, et al.}
\shorttitle{Faint X-ray Bursters}
\title{Sub-luminous X-ray Bursters Unveiled with \integral}
\author{Melania Del Santo}
\affil{INAF/IASF-Roma, via del Fosso del Cavaliere 100, 00133 Roma, Italy}
\author{Lara Sidoli}
\affil{INAF/IASF-Milano, via E. Bassini 15, 20133 Milano, Italy}
\author{Patrizia Romano}
\affil{INAF/IASF-Palermo, via U. La Malfa 153, 90146 Palermo, Italy}
\author{Angela Bazzano, Antonella Tarana, Pietro Ubertini, Memmo Federici}
\affil{INAF/IASF-Roma, via del Fosso del Cavaliere 100, 00133 Roma, Italy}
\author{Sandro Mereghetti}
\affil{INAF/IASF-Milano, via E. Bassini 15, 20133 Milano, Italy}

\begin{abstract}
In 2005 March 22nd, the \integral~ satellite caught a type-I X-ray burst from the unidentified source \src, 
serendipitously discovered with \xmm~in 2003. Based on the type-I X-ray burst properties, 
we derived the distance of the object and suggested that the system is undergoing a prolonged accretion episode of many years. 
We present new data from a \swift/XRT campaign which strengthen this suggestion. 
\ax~was an unclassified source reported in the \asca~catalogue of the Galactic Centre survey. 
\integral~ observed a type-I burst from it in 2005, April 16th. Recently, a \swift~ ToO 
allowed us to refine the source position and establish its persistent nature. 
\end{abstract}

\section{Introduction}
In 2003 the new faint X-ray source \src~ was detected in the Galactic Centre (GC) 
region by \xmm~ at a flux of $(6.8 \pm {0.4}) \times$10$^{-12}$ \ferg~
(Sidoli \& Mereghetti 2003; Sidoli et al. 2004).
In 2005 it was serendipitously re-pointed and detected at a lower flux level 
($\sim 4 \times 10^{-12}$ \ferg), but with the same spectral parameters
($\Gamma$=2.1$\pm${0.1}, \nh=(8.9$\pm${0.5})~$\times10^{22}$~cm$^{-2}$; see Del Santo et al. 2007a).
Brandt et al. (2006) reported on a possible type-I X--ray burst from the new \integral~ source \igr,
that was soon after associated with the transient \src~ \cite{wij06a}.
Analysing the burst properties and the persistent emission from several \xmm~ and \chandra~ observations, 
\src~ was localised in the GC \cite{delsanto07b} and
classified as a ``quasi-persistent'' sub-luminous LMXB with a neutron star \cite{delsanto07a}.  
\swift~ and \chandra~ ToOs strengthened the new findings
that the system has been undergoing a prolonged accretion episode of many years \cite{degenaar07}. 

\ax~ was reported in the \asca~catalogue \cite{sakano02} as an unidentified source 
with unabsorbed $F_{2-10} \sim 1.6 \times 10^{-11}$ \ferg, $\Gamma$=3.7 and \nh $\sim 4.5 \times 10^{22}$ cm$^{-2}$.
An intense photospheric-radius-expansion type-I X-ray burst was detected with JEM-X 
from \ax~ that allowed us to estimate a distance of 6.8$\pm$0.7 kpc (Chelovekov \& Grebenev 2007a).

 \begin{figure}[t!]
  \centering
   \includegraphics[angle=90,height=5cm]{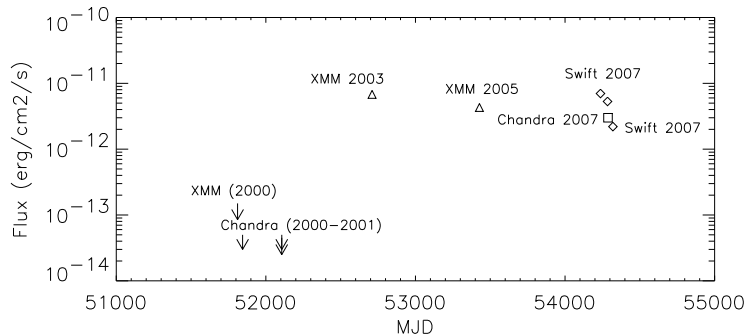}
      \caption{Long term (2-10 keV) lightcurve of \src~ with \xmm\ (triangles),
\chandra\ (square) and \swift\ (diamonds). Upper limits are reported with arrows.}
\label{fig:lc_xmm}
   \end{figure}
 
\section{Observations and Results}
\swift/XRT observed \src~ on 13 and 17 May 2007 (Degenaar \& Wijnands 2007),
and on 3 July 2007 and on 8 August 2007 \cite{sidoli07} as part of our \swift~ ToO campaign of the source. 
Here we report the results on the latter observations.

\ax~ was observed with \swift/XRT as a 3 ks Target of Opportunity on July 12 2007 \cite{delsanto07c}.
We also present some \integral/JEM-X results of the \ax~ X-ray burst event 
which occurred on 2005 April 16. A detailed analysis has been presented in Chelovekov \& Grebenev (2007b).

\src~ has been detected with a 0.2--10 keV count rate of $(3.56\pm 0.71) \times 10^{-2}$ counts s$^{-1}$ (3 July 2007) 
and  $(1.50\pm 0.35) \times 10^{-2}$ counts s$^{-1}$ (8 Aug 2007), which translates into an unabsorbed 2-10 keV flux
of $(5.3 \pm 1.0) \times 10^{-12}$ \ferg~ and $(2.2 \pm 0.5)\times 10^{-12}$ \ferg.
Because of the poor statistics, an absorbed power law model with $\Gamma$=2.25 and \nh=9$\times 10^{22}$ cm$^{-2}$
has been assumed.
A comparison of these values with the previous measurements started in 2003 (reported in Del Santo et al. 2007a)
reveals a complex lightcurve (Fig. \ref{fig:lc_xmm}), 
apparently not compatible with a simple linear decay. 
Nevertheless, the outburst seems to now be in a declining phase.   

\begin{figure*}[t!]
 \centering
   \includegraphics[height=4.6cm]{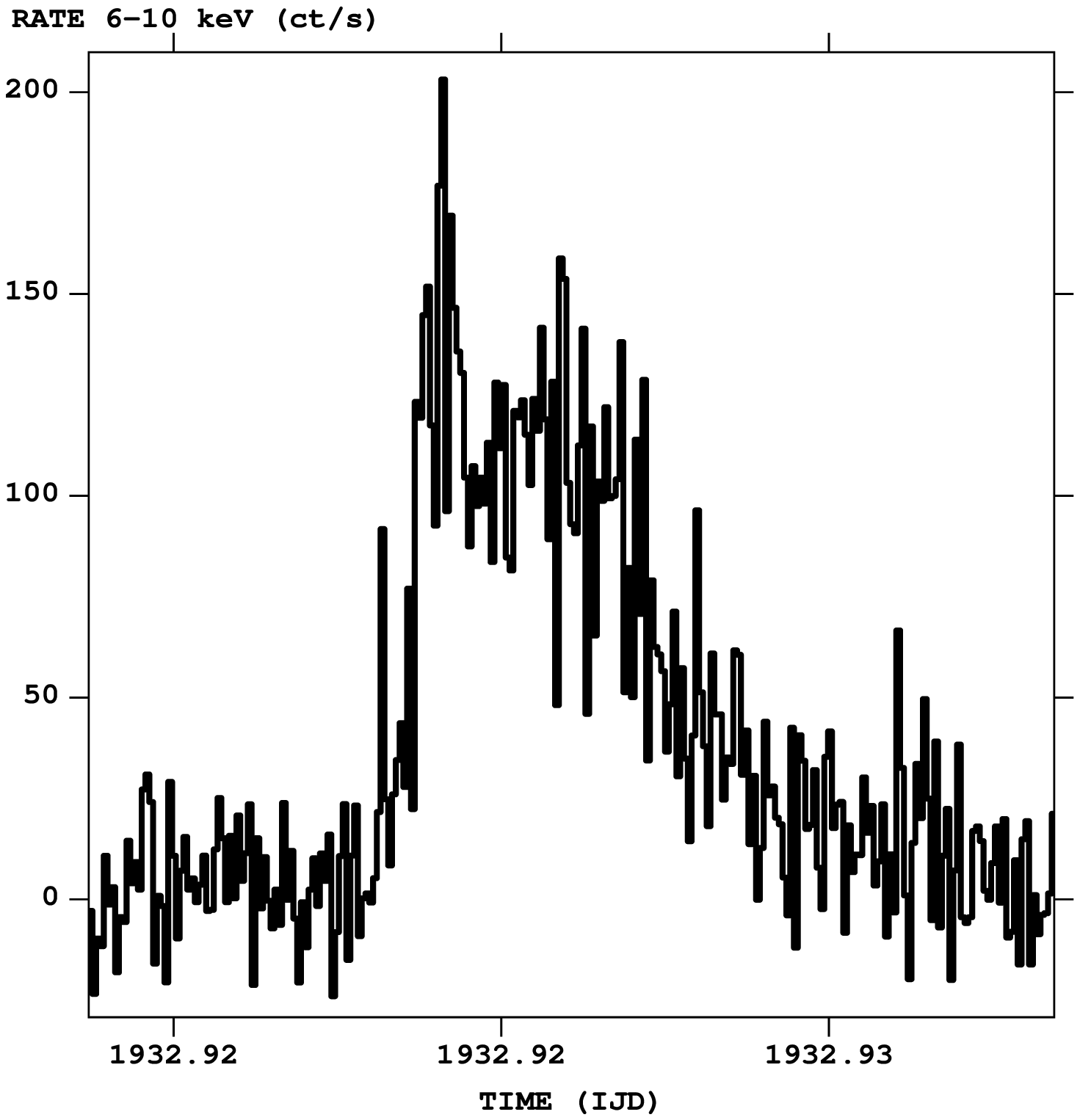}
 \includegraphics[height=4.3cm]{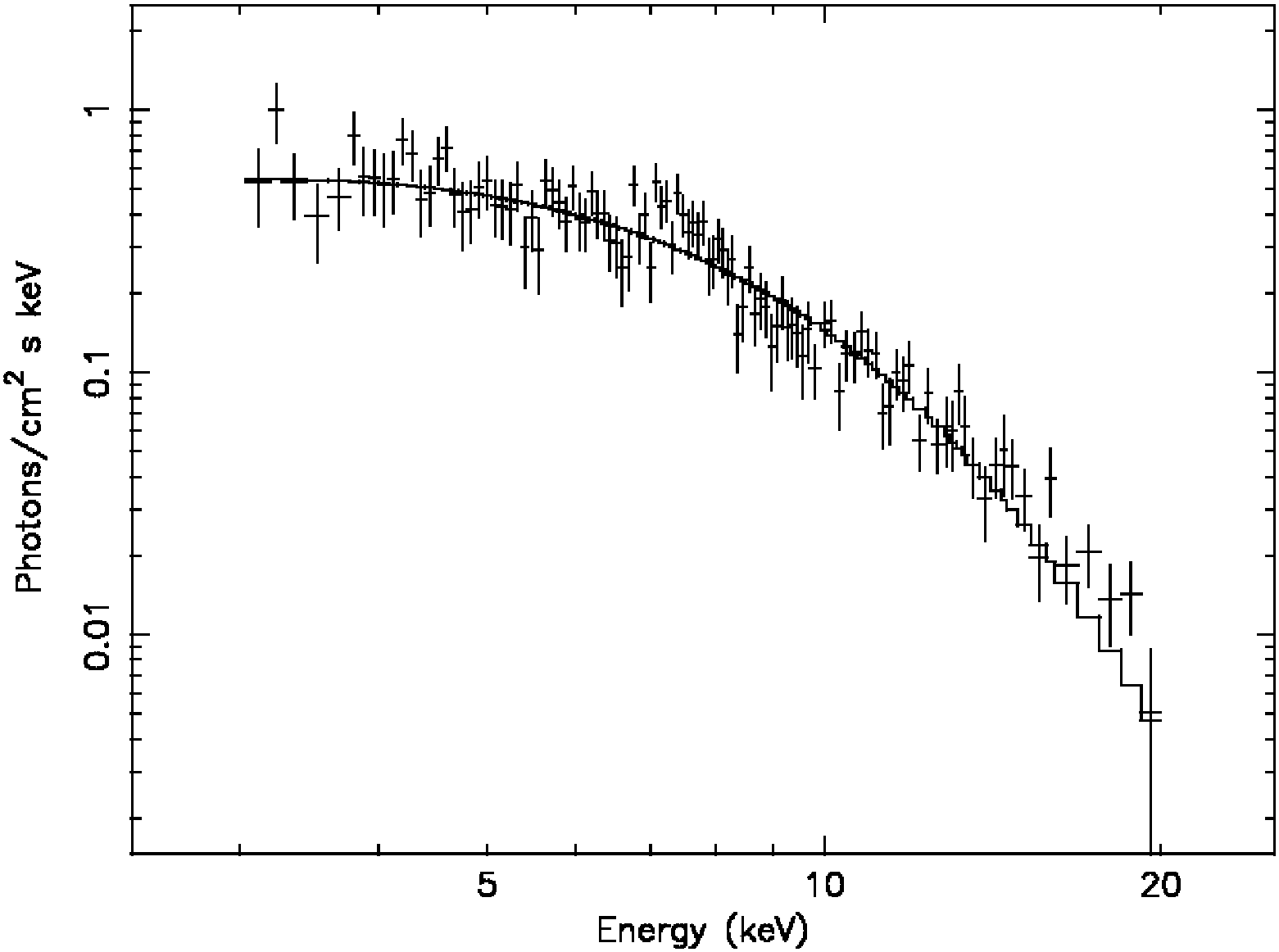}
 \includegraphics[height=4.5cm]{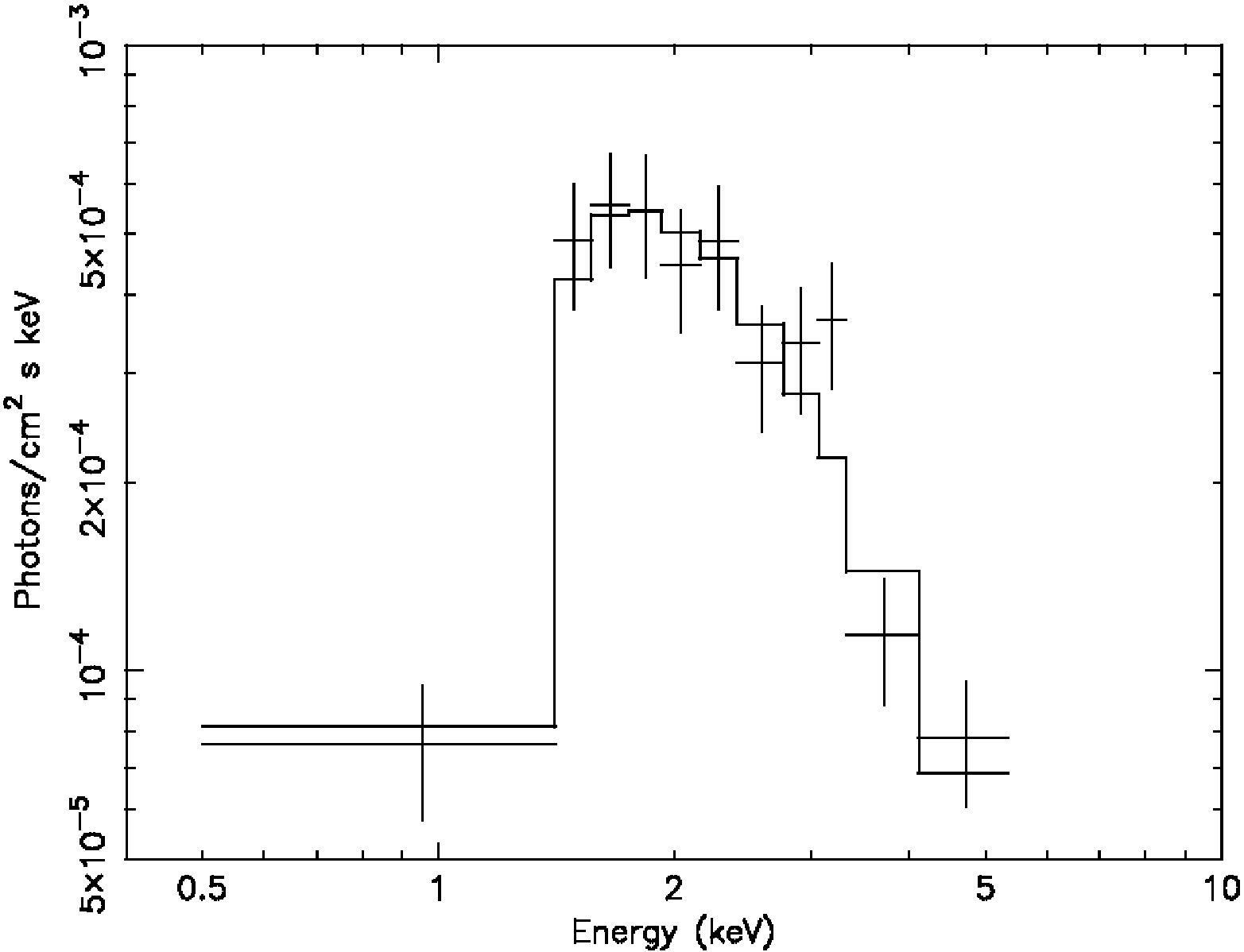}
     \caption{\ax~ X-ray burst light-curve in the 6-10 keV band observed by JEM-X ({\it left}) 
and black-body spectrum ({\it centre}) 
are shown. On the {\it right} the \swift/XRT photon spectrum of the persistent emission is shown.}
\label{fig:ax}
   \end{figure*}

The \ax~ type-I X-ray burst showed a double-peak profile (Fig. \ref{fig:ax}, {\it left})
with a peak flux of about $5.7 \times 10^{-8}$ \ferg~ and a black-body
spectrum with $kT_{bb}=1.96 \pm 0.06$ keV (Fig. \ref{fig:ax}, {\it centre}). 
Thanks to \swift/XRT, we could observe the persistent emission of the source and provide the refined position as 
RA(J2000)=17h 54m 14.6s and Dec(J2000)=-27d 54m 34.3s \cite{delsanto07c}. 
The best spectral fit is an absorbed power-law with $\Gamma$=$3.6 \pm 0.7$ and \nh=$(2.8 \pm 0.7) \times 10^{22}$ 
cm$^{-2}$ (Fig. \ref{fig:ax}, {\it right}). The unabsorbed flux in the 2-10 keV band is $6.3 \times 10^{-12}$ \ferg. 
Assuming a distance of 6.8 kpc \cite{chelo07a} this translates into a luminosity of 
$4 \times 10^{34}$ \lerg, which is more than a factor of 2 lower than the one observed in 1999 by \asca~ ($\sim 10^{35}$ \lerg). 

\section{Conclusions}
The \integral~ discovery of type-I X-ray bursts from \src~ and \ax~ allowed us to establish the nature of 
sub-luminous ($L_{X}=10^{34}-10^{35}$ \lerg) cneutron stars in low mass systems and to evaluate their distances.
Thanks to the X-ray observations of the persistent emission, \src~ could be classified as a ``quasi-persistent''
very faint X-ray transient, while \ax~ is likely a persistent source.
In the \sax~ era \src~ and \ax~  would have been included in the ``burst-only'' sources list \cite{corne04}, 
because of the WFC limit sensitivity (10 mCrab, $10^{36}$ \lerg in the GC).
In fact they are not present in the WFC catalogue \cite{capitanio08}.
There are fewer X-ray bursters in the lowest accretion regime than predicted \cite{corne04}.
\src~ and \ax~ increase the number of bursters observed  at $\dot{M} < 10^{-10}$  M$_{\odot}$/yr.

\begin{acknowledgements}
\footnotesize{
Based on observations obtained with \xmm\ and \integral, ESA science
missions and \swift, NASA science mission. 
Data analysis is supported by the Italian Space Agency (ASI),
via contract ASI/INAF I/023/05/0 and ASI/INTEGRAL I/008/07/0.
We thank N. Gehrels for accepting our \swift~ToO proposal and J. Chenevez and S. Brandt for JEM-X analysis discussion.}
\end{acknowledgements}

\end{document}